\documentstyle[prl,aps]{revtex}
\begin{document}
\preprint{SU-ITP-94-6, cond-mat/9403038}

\title{ Fluctuation-Induced First Order
Transition\\ between the Quantum Hall Liquid and Insulator}
\author{Leonid Pryadko and Shou-Cheng Zhang}
\address{ Department of Physics, Stanford University, Palo Alto, CA 94305}
\date{March 9, 1994}
\maketitle
\begin{abstract}
We study the phase transition between the quantum Hall liquid state
and the insulating state within the framework of the
Chern-Simons-Landau-Ginzburg theory of the quantum Hall
effect. For the transition
induced by a background periodic potential in the absence of disorder,
the model is described by a relativistic scalar field coupled to the
Chern-Simons gauge field. For this system,
we show that the transition is of the first order,
induced by the fluctuations of the gauge field, rather than second
order, with statistical angle-dependent scaling exponent.
\end{abstract}

Recently the Chern-Simons-Landau-Ginzburg (CSLG) theory~\cite{zhk,zhang}
of the quantum Hall effect
has been applied to study the phase transition between the quantum Hall
liquids and between quantum Hall liquids and insulators~\cite{global}.
The central idea of this theory is to map the transition in the quantum
Hall systems to the transition from superfluid to insulator. For the primary
quantum Hall states, the off-diagonal-long-range-order of the quantum
Hall liquid state~\cite{girvin} allows one to
identify it with the superfluid state of the boson, and the insulating state
in the quantum Hall systems, the Hall insulator, is identified with the
insulating state of the bosons. The resulting theory describes
non-relativistic bosons coupled to the Chern-Simons gauge field. The
transition between the superfluid and the insulating state of the boson is
induced either by disorder or the magnetic field. Using the CSLG one can
derive a set of ``laws of corresponding states''~\cite{jain,global}
relating the
disorder-induced second order phase transition in the integer quantum Hall
effect to transitions in the fractional quantum Hall regime, as long
as these transitions are induced by disorder as well.
Using this theory, Kivelson, Lee and Zhang~\cite{global}
constructed a global phase diagram of the quantum Hall effect in the two
dimensional parameter space of disorder versus magnetic field and identified
relations between the various inter-plateau transitions and transition
between quantum Hall liquid and the Hall insulator.
In the case of strong disorder, the integer transition is widely believed
to be of the second order. From the point of view of the law of
correspondence,
the fractional transition in this limit should be of the second order
as well~\cite{wang}. This point of view is supported by the experiment
of Wei {\it et al}~\cite{wei} where similar critical exponents were measured
for both integer and fractional transition.
However, the situation is less clear when the transition is
dominated by interaction rather than disorder. There are some theoretical
indication that the transition from a quantum Hall liquid to the Wigner
crystal is of the first order~\cite{wigner}.
In this paper we shall study the order of the transition in the
absence of disorder within the framework of CSLG theory.

Within the CSLG theory, the transition in the quantum Hall effect without
disorder is mapped to a boson superfluid to insulator transition driven
either by interaction or some periodic potential.
It is argued that the transition boson superfluid to insulator
induced by background potential belongs to the same universality class of
three dimensional XY model, or relativistic scalar field theory with a mass
term and $\phi^4$ interaction~\cite{fisher}.
Applying CSLG theory to this system naturally leads to a model of
relativistic scalar field coupled with the Chern-Simons term~\cite{wenwu}.
The critical behavior is obtained by tuning the mass of the scalar field to
zero. Wen and Wu~\cite{wenwu} studied this model in the limit where $N$, the
number of components of the scalar field, is large. They found
that the Chern-Simons gauge field is a marginal
perturbation to the scalar field fixed point, so that the critical exponents
depend on the coefficient of the Chern-Simons term, or the statistical angle
that is nothing but the filling fraction of the quantum Hall
states~\cite{zhk,zhang}.  The same phenomenon was found in the Mott
transition of anyons on a lattice~\cite{mott}, where the scaling exponents
depend on the statistics of the anyons.

In this paper we show that in the physically relevant case of $N=1$,
the phenomenon of statistical
angle-dependent critical exponents does not occur in this model.
In fact there is no critical point in the
weak-interacting theory when gauge field is present! Instead,
the transition is
of the first order, driven by the fluctuations of the gauge field.
This is a well known phenomenon in both particle physics~\cite{coleman} and
condensed matter physics~\cite{halperin}.  Coleman and Weinberg~\cite{coleman}
studied the four-dimensional abelian Higgs model and found that
even if one tunes bare mass to zero, the fluctuations of the gauge field
always generate a mass dynamically. Thus there is no critical point in the
theory.  This effect goes is called ``dynamical mass generation'' or
``dimensional transmutation.''  Independently, Halperin, Lubansky and
Ma~\cite{halperin} discovered the same phenomenon in the Ginzburg-Landau
theory of superconductor to normal metal transition, and showed that the
fluctuations of the electromagnetic field induce a first order transition.
Our work is basically an application of these ideas to the quantum Hall
systems. In view of the fact that the transition is described by the (three
dimensional) field theory of a scalar coupled to the gauge field, it is not
surprising the phenomenon of fluctuation-induced first order phase
transition also occurs here.

Since the CSLG model is constructed from some general principles, we
believe that our analysis implies that the transition from a quantum Hall
liquid to an insulating state in the absence of disorder is generally a first
order transition. This model can be useful in understanding the transition
from the quantum Hall liquid state to the Wigner crystal state which is
believed to be a first order transition~\cite{wigner}.  Since there is
no critical point in this model, it can not be used to address any
questions about the critical properties of the disorder-induced
transition in the quantum Hall systems. In particular, the
phenomenon of statistical-angle-dependent critical exponents
does not occur in this model, therefore there is no theoretical
basis to conjecture that similar phenomenon would occur in
disorder-driven transition in the quantum Hall effect~\cite{wenwu}.

The Lagrangian of interest is the 2+1 dimensional relativistic scalar field
coupled to the Chern-Simons gauge field. In Euclidean space it is given by
\begin{eqnarray}\label{Lagrangian}
{\cal L}[a,\phi]=\sigma_{xy}\frac{i}{2}\epsilon^{ijl}a_{i}
\partial_{j}a_{l}
+\frac{1}{2}\left|(\partial_j+ia_j)\phi\right|^2\\
+\frac{1}{2}m^2|\phi|^2+\frac{\lambda}{2}|\phi|^4
+\frac{g}{3}|\phi|^6\nonumber
\end{eqnarray}
where $\sigma_{xy}=1/\theta\equiv 1/2\pi q$ is the Hall conductivity,
$q$ is an odd integer,
and  $m$, $\lambda$ and $g$ are the mass and interaction constants of the
scalar field. We have explicitly included the sixth
order term, which is a marginal operator in $3$ dimensions. A
logarithmically divergent contribution to the sixth order term occurs in
the loop expansion, therefore, keeping this term is necessary
to ensure renormalizability.

When the complex field $\phi$ develops a vacuum expectation value, the
$U(1)$ symmetry is spontaneously broken. This superfluid phase of the
boson field corresponds to the quantum Hall liquid state~\cite{zhang}.
On the other hand, the insulating state corresponds to the case
when the vacuum expectation value of the field $\phi$ vanishes. At the
level of the classical potential, this transition occurs at $m=0$.
The $\lambda \phi^4$ interaction is a relevant perturbation to this
Gaussian fixed point. The associated infrared divergences are usually
controlled either by introducing a large $N$ generalization of the model and
carrying out a systematic $1/N$ expansion, or by staying close to the
critical dimension of four and carrying out an $\epsilon = 4-D$ expansion.
In the present context, however, the $1/N$ expansion
has the drawback of underestimating the effect of the gauge field
fluctuations. Since there are $N$ scalar components and only one component
of the gauge field, the effect responsible for fluctuation-induced first
order phase transition is not visible to the first order in $1/N$.
For example, in the case of a superconductor to normal state transition,
the gauge fluctuations only induce a first order transition when $N<365.9$
\cite{halperin}. On the other hand, the $\epsilon$ expansion from four
dimensions is not well suited for the Chern-Simons term which is naturally
defined at $D=3$. In this work we control the infrared
divergences associated with the massless point $m=0$ by staying close
to the ``tricritical point''~\cite{tri}
where $\lambda=0$ as well, and studying the effects of
marginal operators $\sigma_{xy}$ and $g$ within the loop expansion.

Without the coupling to the Chern-Simons gauge field, the point
$m=\lambda=0$ is tricritical: for a finite positive $g$
the transition is of the first order when $\lambda<0$ and it is of the second
order when $\lambda>0$; the first and the second order lines meet at the
tricritical point.
The critical dimension for the theory at the tricritical
point is $D=3$, one can therefore carry out a systematical perturbation
expansion in the number of loops.
We shall show that with the coupling to the Chern-Simons
gauge field, the physics is changed fundamentally:
while the model at the tricritical point is massless and has no
dimensionful coupling constants at the classical level, quantum
fluctuations associated with the Chern-Simons gauge field induce an
effective potential which has a minimum away from $\phi=0$ and
some dynamically generated mass scale. Since the physical mass scale at the
point $m=\lambda=0$ is finite, introducing a small finite
$\lambda$ could not lead to uncontrolled infrared divergences. From this
argument one can show that the fluctuation-induced first order transition
may occur even in the presence of a finite $\lambda$.

We first present our result of one loop effective potential calculated within
the path integral technique.  Within one-loop accuracy, we assume a
space-independent configuration of the scalar field $\phi(x)=\phi_0$, and
integrate out the fluctuations due to both the scalar and the Chern-Simons
gauge fields. The equivalent sequence of Feynman diagrams for this
calculation is shown in Figure 1.

At the classical level,
the Chern-Simons gauge propagator is purely off-diagonal; it is proportional
to $\epsilon^{ijl} \hat{p}_l/p$. At the quantum level, however, a finite
renormalization to the Chern-Simons propagator is induced.
The general form of ``dressed'' propagator, accounting for this correction,
is given by
 \begin{equation} D_{ij}=\rho_{xx}\frac{p^2\delta_{ij}-p_i p_j}{p^3}-
        \rho_{xy}\frac{\epsilon^{ijl}p_l}{p^2},
\label{GaugePropagator}
\end{equation}
where, as usual,
$\rho_{xx}=\sigma_{xx}/(\sigma_{xx}^2+\sigma_{xy}^2),$
  $\rho_{xy}=\sigma_{xy}/(\sigma_{xx}^2+\sigma_{xy}^2).$
At the one loop level $\sigma_{xy}$ is not renormalized, whereas a finite
$\sigma_{xx}=N/24$ is obtained for the $N$-component scalar field.
We shall use this form of the gauge propagator to keep our analysis general.

The one-loop effective action is defined by
\begin{equation}
S_{\it eff}[\phi_0]=S[0,\phi_0]+\frac{1}{2}\ln\det
\left(\left.\frac{\delta^2 S[a,\phi]}
{\delta [a,\phi]^2}\right|_
{\stackrel{\!\!\!\!\scriptstyle a=0}{\scriptstyle \phi=\phi_0}}
\right),
\label{Determinant}
\end{equation}
where the Euclidean action is
$S[a,\phi]=\int {d^3x {\cal L}[a,\phi]}.$
We calculate this fluctuation determinant at $\phi_0=\rm const$
within the covariant gauge
$\partial_i a_i=0$ by the $\zeta$-function regularization scheme.
The result is given by
\begin{eqnarray}
&\relax&\delta V(\phi)=\frac{4\rho_{xx}(3\rho_{xy}^2-
\rho_{xx}^2)\phi^6}{3\pi^2}\left(\ln\frac{2\phi^2}{M}
-\frac{11}{6}\right)\\
&\relax&-\frac{\left(
\left(m^2+3\lambda\phi^2+3g\phi^4\right)^{3/2}+
\left(m^2+\lambda\phi^2+g\phi^4\right)^{3/2}\right)}{12\pi},
\nonumber
\end{eqnarray}
where $M$ is a mass scale introduced by the regularization procedure.
Note, that within the first loop approximation regularized corrections
due to self-interactions contain no logarithmic divergences.

Next we introduce normalization conditions to define values of the
renormalized coupling constants. Because of the logarithmic dependence
of our effective potential on $\phi$, the coupling constant $g$ has to be
defined at some finite scale $\mu$
	\begin{equation}V'''_{\phi^2}(\phi^2=\mu)=2g.
	\label{normalization}\end{equation}
Also, in the tricritical point
	\begin{equation}
	V''_{\phi^2}(\phi^2=0)=\lambda=0,\quad
	V'_{\phi^2}(\phi^2=0)=m^2=0.
	\label{normalization2}\end{equation}
With these normalization conditions, we
arrive at the one-loop effective potential
\begin{equation}
V_{\it eff}(\phi)=\left(\!g\!+\!\frac{4\rho_{xx}(3\rho_{xy}^2\!-\!
\rho_{xx}^2)}{\pi^2}
        \left(\!\ln{\frac{\phi^2}{\mu}}
-\frac{11}{6}\!\right)\right)\frac{\phi^6}{3}.
\label{EffectivePotential}
\end{equation}
The qualitative features of this effective potential are very similar to
those obtained from the scalar electrodynamics in
$4$~dimensions~\cite{coleman}. In four
dimensions, both coupling constants, the electric charge $e$ and
$\lambda$ are dimensionless and the resulting effective potential
can be viewed as logarithmic corrections to the $\lambda \phi^4$ coupling.
Here in $3$ dimensions, both the Chern-Simons coefficients, $\rho_{xx}$
and $\rho_{xy}$, and the coupling $g$ are dimensionless; the resulting
effective potential can be viewed as logarithmic corrections due to these
marginal operators. The effective potential is bounded when
$3\rho_{xy}^2>\rho_{xx}^2$, which is the clean fractional
quantum Hall system we are interested in.
In the opposite case, the effective potential is unbounded
at the one loop level, however, we believe that higher order corrections will
cure this unphysical feature~\cite{lubensky}. In the case of
$3\rho_{xy}^2>\rho_{xx}^2$ one can easily see that
(\ref{EffectivePotential}) has a global minimum at $\phi^2_{\it
min}=\mu\exp(3/2-g/\alpha),$ away from the origin, and the value of the
effective potential at this minimum $V_{\it eff} (\phi_{\it
min})=-(5\alpha\mu^3/6)\exp(9/2-3g/\alpha)$
is negative. Here $\alpha$ is the numerical coefficient in front of the
logarithmic term in (\ref{EffectivePotential}). Furthermore, the effective
mass at this new minimum is given by $m_{\it eff}^2=V_{\it
eff}''(\phi_{min})/2=2\alpha\mu^2 \exp(3-2g/\alpha)$.  This is the general
feature of the dimensional transmutation: although the classical model at the
tricritical point $m=\lambda=0$ is free of any dimensionful parameters,
a mass scale is generated dynamically by the quantum fluctuations.
The classical tricritical point is in fact not a critical point at all!

Our analysis is carried out at the point $m=\lambda=0$. Without the
coupling to the Chern-Simons gauge field, this is a Gaussian fixed point
if $g=0$ and a tricritical point if $g>0$. In both cases $\lambda$ is a
relevant perturbation in three dimensions. Conclusions reached for the
case $\lambda=0$ certainly do not carry over to the case of finite
$\lambda$ because of the strong infrared divergences associated with this
relevant perturbation.
In the present case, however, since the effective mass at
the point $m=\lambda=0$ is finite, there is a finite ``Ginzburg
region" defined by
        \begin{equation}
        \lambda  \ll m_{\it eff},    \qquad    m^2
        \ll m_{\it eff}^2.\label{Conditions}
        \end{equation}
As long as $\lambda$ and $m$ lie inside this region, their
effects can be treated classically.
 By studying
the structure of the minimum of the total potential,
$\frac{1}{2}m^2|\phi|^2+\frac{\lambda}{2}|\phi|^4+V_{\it eff},$
we obtain the phase diagram shown in Figure 2. It is plotted in
the scaled variables, $m_*^2=(m^2/\mu^2\alpha)\exp(2 g/\alpha)$ and
$\lambda_*=(\lambda/\mu\alpha)\exp(g/\alpha).$
Note, that while the
position of the tricritical point is shifted, it is still far away
from the edge of the
applicability of the theory which in the dimensionless
variables is given by
        $\lambda_*\ll\sqrt{2}e^{3/2},\quad
                m_*^2\ll 2 e^3;$
this implies that the fluctuations due to $\lambda$ and $m$ do
not change the
qualitative features of the phase diagram. From Figure 2 we see
explicitly that the transition is of the first order for small enough
$\lambda$.  For large enough $\lambda$ the transition from the insulator to
quantum Hall liquid may be of the second order, and there is also a first
order transition between two quantum Hall liquid states. However, one may
have to go to higher orders in perturbation theory to fully understand
transitions in these regions.

A drawback of the one-loop approximation
is that logarithmic corrections to
the potential $g\phi^6$  vanish when $\rho_{xx}=0$, therefore, if we
strictly use the one loop approximation with the pure Chern-Simons
propagator rather than equation (\ref{GaugePropagator}), the regularized
effective potential do not change. It turns out that
at the second loop level the
finite renormalization to the gauge propagator is automatically included.
Therefore, to make our analysis complete, we have carried out the two loop
calculation of the effective potential. This amounts to a calculation of the
$\beta$ function for $g$ and the wave function renormalization factor
$\gamma$. Also, we have checked that infinite corrections for $\rho_{xx}$
and $\rho_{xy}$ vanish in the second loop order (this result for the pure
Chern-Simons propagator has been obtained earlier~\cite{ExactTwoLoop}.)
The effective potential is obtained from the solution of the
renormalization-group (RG) equation:
\begin{equation}
\left(\mu\frac{\partial}{\partial\mu}+
\beta(g)\frac{\partial}{\partial g}+ 2\gamma(g)
\phi^2\frac{\partial}{\partial\phi^2}\right)V=0,
\label{Callen}
\end{equation}
subject to initial conditions
(\ref{normalization},\ref{normalization2}).
To the second loop order the coefficients are
\begin{eqnarray}
2\gamma&=&
-\left(\frac{4\rho_{xx}}{3\pi^2}
+\frac{7\rho_{xy}^2}{24\pi^2} \right),
\label{gamma}\\
\label{beta}
\beta&=&
-\frac{12}{\pi^2}\rho_{xx}\rho_{xy}^2
-\frac{33}{4\pi^2}\rho_{xy}^4\\
&\quad &+\, g\left(\frac{37}{8\pi^2}\rho_{xy}^2
+\frac{4}{\pi^2}\rho_{xx}\right)
-\frac{7}{12\pi^2}g^2.\nonumber
\end{eqnarray}
 From (\ref{beta}) we see explicitly that $g$ flows to negative values if
its bare value is small enough,
        $$g<g_*\approx3\rho_{xy}^2
        \frac{\rho_{xx}+11\rho_{xy}^2/16}
                     {\rho_{xx}+37\rho_{xy}^2/32}.
    $$
This ``run-away'' trajectory is usually taken as an indication of the
first order phase transition~\cite{halperin}.
For small $g$, we can keep only
the first two terms in the expansion of the $\beta$-function in $g,$
$\beta(g)= \beta_0+\beta_1 g.$ Integrating the RG equation within
this approximation we obtain the effective potential
\begin{equation}
V(\phi)=2\phi^6\left(\!\frac{g-g_*}{(3+\delta)(2+\delta)(1+\delta)}
\left(\frac{\phi^2}{\mu}\right)^{+\delta}\!\!\!\!+\frac{g_*}{6} \right)\!,
\label{TwoLoop}
\end{equation}
where
        $\delta=\beta_1/(1-2\gamma)\approx\beta_1.$
The second-loop order effective potential (\ref{TwoLoop}) has
a global minimum away from the origin, indicating the first
order transition. In our approximation it is bounded for all
physical values of $\rho_{xx}$ and $\rho_{xy}$ and the first
order transition occurs even when $\rho_{xx}=0$.

In conclusion, we have investigated the order of the transition from a
quantum Hall liquid state to an insulating state {\it in the absence
of disorder}. Within the CSLG theory, we showed that this transition
in generally of the first order, induced by the fluctuations of the
Chern-Simons gauge field. Because there is no critical point in
this theory, the
phenomenon of the statistical-angle-dependent critical exponents
does not occur.
We believe that our theory is generally applicable to quantum Hall phase
transitions dominated by interaction rather than disorder, which is the
opposite limit compared to that studied in the work on the global
phase diagram~\cite{zhang}. In
particular, this theory could provide a framework to study the transition
from a quantum Hall liquid to a Wigner crystal in the limit of pure samples.

We would like to thank S. Kivelson for a stimulating discussion that motivated
this work and R.~Kallosh for discussion on some aspects of
perturbative expansions in the quantum field theory.
S.~C.~Z. is supported in part by the Center for Materials Research at
Stanford University and L.~P. would like to thank the IBM Corporation for
support through the IBM graduate fellowship program.

\newpage

\begin{figure}[p]
     \caption{Gauge loops contributing to the effective
action ({\protect\ref{EffectivePotential}}).}
\label{FirstLoop}
\end{figure}

\begin{figure}[p]
     \caption{The phase diagram of the one-loop effective
action ({\protect\ref{EffectivePotential}}) in the scaled coordinates
        $ m_*^2={m^2}e^{2g/\alpha}/{\alpha\mu^2}$ versus
        $\lambda_*={\lambda}e^{g/\alpha}/{\alpha\mu}.$
The solid line
represents the first order phase transition between the non-symmetric and
symmetric phases {\bf A} and {\bf B} for $m_*^2>0;$ for $m_*^2<0$
it is the first order transition between two non-symmetric phases {\bf A} and
{\bf C} with different expectation values of the order parameter $\phi.$ The
second order transition between {\bf B} and {\bf C} is shown with the dashed
line.  Captions draft the effective potential $V_{\it eff}(\phi)$ in
appropriate regions. Dots show the phase boundary without the coulping to
the gauge field.}
\label{PhaseDiagram}
\end{figure}
\end{document}